\documentclass[onecolumn,showpacs,preprintnumbers,amsmath,amssymb,preprint]{revtex4}
\usepackage{graphicx}
\usepackage{dcolumn}
\usepackage{bm}
\begin{document}


\title{Analytical determination of  the stop band tuning of 
photonic crystals infiltrated with liquid crystals}

\author{J. Manzanares-Martinez, P. Castro-Garay, E. Urrutia-Ba\~nuelos,
D. Moctezuma-Enriquez, R. Archuleta-Garcia and M. A. Velarde-Chong}
\affiliation{Departamento de Investigaci\'on en F\'isica de la Universidad
de Sonora, Apartado Postal 5-088, Hermosillo, Sonora 83190,
M\'exico}

\date{\today}
\begin{abstract}
We demonstrate that the tuning of the optical properties of a 
photonic crystal infiltrated with liquid crystal 
can be calculate using the Von-Laue diffraction condition.  We present a
simple formula to predict the shift of the stop band for all
the diffraction orders  using  an effective index of the
composite structure. We consider that our formula is useful to
determine in a simple manner the shift of the optical properties of 
tunable photonic crystals. We compare the accuracy of our method with calculations 
obtained with the Plane Wave Method.
\end{abstract}
\pacs{Valid PACS appear here}

\maketitle

\section{Introduction}
Photonic crystals (PC) are a new kind of materials which facilitate the
control  of the light. PC exhibits Photonic Band Gaps (PBG) that 
forbids the  radiation propagation  in a specific range of frequencies. \cite{Yablonovitch_1987} In the
past ten years has been developed an intensive effort to study and  micro-fabricate PBG
materials in one, two or three dimensions. \cite{Joannopoulos,Guida_2003,Bush_2006}

However, for many applications it is advantageous to obtain some degree of
tuning of the PBG. \cite{Bush_2006} This can be made by 
changing the refractive index of the constitutive materials by means of  an   external parameter,
such the temperature or voltage. \cite{Manza2,Manza3} 
One of the most promising routes to achieve such tuning is the use of 
liquid crystals (LC).\cite{bush_1999}
Recently it  has  been reported several works of PC infilled with LC with robust 
 ranges of tuning in the optical regime. \cite{Tolmachev_2007a,
Tolmachev_2008b,Jim_2008,Arriaga_2008,Dorjgotov,Bush1}
In most of the cases,
the theoretical analysis is performed via the calculation
of the band structure or light reflection (transmission) using plane wave
expansion or Transfer Matrix Methods (TMM).\cite{Jim_2008,Arriaga_2008} In this work we illustrate  
the possibility  of  determining the degree of
tuning in a simple manner by using the Von-Laue condition (VLC). The VLC 
takes account of the variation of the constitutive parameters of the unitary
cell introducing an effective refractive index that averages the value of the
constitutive parameters. This formula is useful to determine in a easy way the
tuning of the recently reported experimental 
configurations. \cite{Tolmachev_2007a,Tolmachev_2008b,Jim_2008,Arriaga_2008,Dorjgotov,Bush1}
The performance and limits of our approximation is compared with exact calculations
obtained with the Plane Wave Method (PWM).\cite{Archuleta_2007}

\begin{figure}[h]
\centerline{\includegraphics[width=0.15\columnwidth,draft=false]{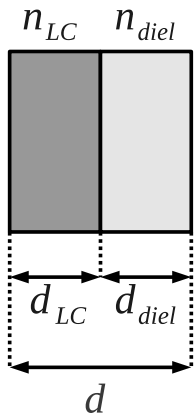}}
\caption{Unit cell of an one dimensional photonic crystal composed by 
liquid crystal and dielectric. The  refractive indices are 
$n_{LC}$ and $n_{diel}$, respectively. The width of each slab is $d_{LC}$ and $d_{diel}$.
The width of the unit cell is $d=d_{LC}+d_{diel}$}
\end{figure}

\begin{figure}[h]
\centerline{\includegraphics[width=1.0 \columnwidth,draft=false]{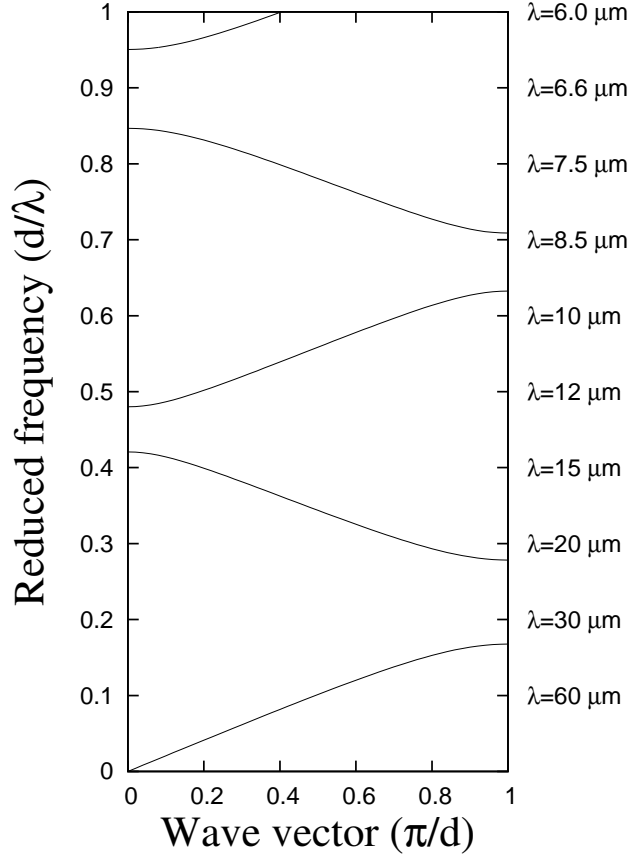}}
\caption{Photonic band structure of a PC composed of a dielectric ($n_{diel}=3.42$) and 
LC ($n_{LC}=1.49$). The filling fraction is $f=0.4.$}
\end{figure}

\section{Theory}

We start our analysis by considering the Von-Laue diffraction condition, \cite{Kittel}

\begin{equation}
\mathbf{k}^2=(\mathbf{k}+\mathbf{G})^2.
\end{equation}

 $\mathbf{k}$ and $\mathbf{G}$ are the wave vector and reciprocal 
lattice vector, respectively. If the vectors $\mathbf{k}$ and $\mathbf{G}$ are parallels, it can be written

\begin{equation}
2 |\mathbf{k}| = |\mathbf{G}|,
\end{equation}

where the absolute value of the reciprocal latttice vector ($|\mathbf{G}|$) is defined as 

\begin{equation}
|\mathbf{G}|= \frac{2 \pi}{d} m,
\end{equation}

$d$ and $m$ are the lattice period and the diffraction order, respectively. 
The wave vector $|\mathbf{k}|$ is defined as

\begin{equation}
|\mathbf{k}| = n_{eff} \frac{\omega}{c}, 
\end{equation}

where $n_{eff}$ is the effective index of the composite medium. 
Using eqs. (3), (4) and (2)  the VLC is

\begin{equation}
\frac{d}{\lambda_m} = \frac{m}{2 n_{eff}},
\end{equation}

where we have introduced the definiton of reduced frequency $\omega d/2\pi c={d}/{\lambda_m}$, 
with $\lambda_m$ as the wavelength for the diffraction order $m$. 
Fig. 1 shows a unit cell composed  of liquid crystals and dielectric, with refractive indices
 $n_{LC}$ and $n_{diel}$, respectively. The width of each layer
is $d_{LC}$ and $d_{diel}$. The width of the unit cell is  $d=d_{LC}+d_{diel}$.
We define the filling fraction  as the space filled by the dielectric material
over the total space in the unit cell, $f=d_{diel}/d$.  
The  effective index $n_{eff}$ is  taken as a simple  average in the unit cell in the form

\begin{equation}
n_{eff} = f n_{LC}  + (1-f) n_{diel}
\end{equation}

The tuning of the diffraction condition can be written as

\begin{equation}
\frac{d}{\lambda_m} = \frac{m}{2[f n_{LC}  + (1-f) n_{diel}]}, 
\end{equation}

\begin{figure}[h]
\centerline{\includegraphics[width=1.0 \columnwidth,draft=false]{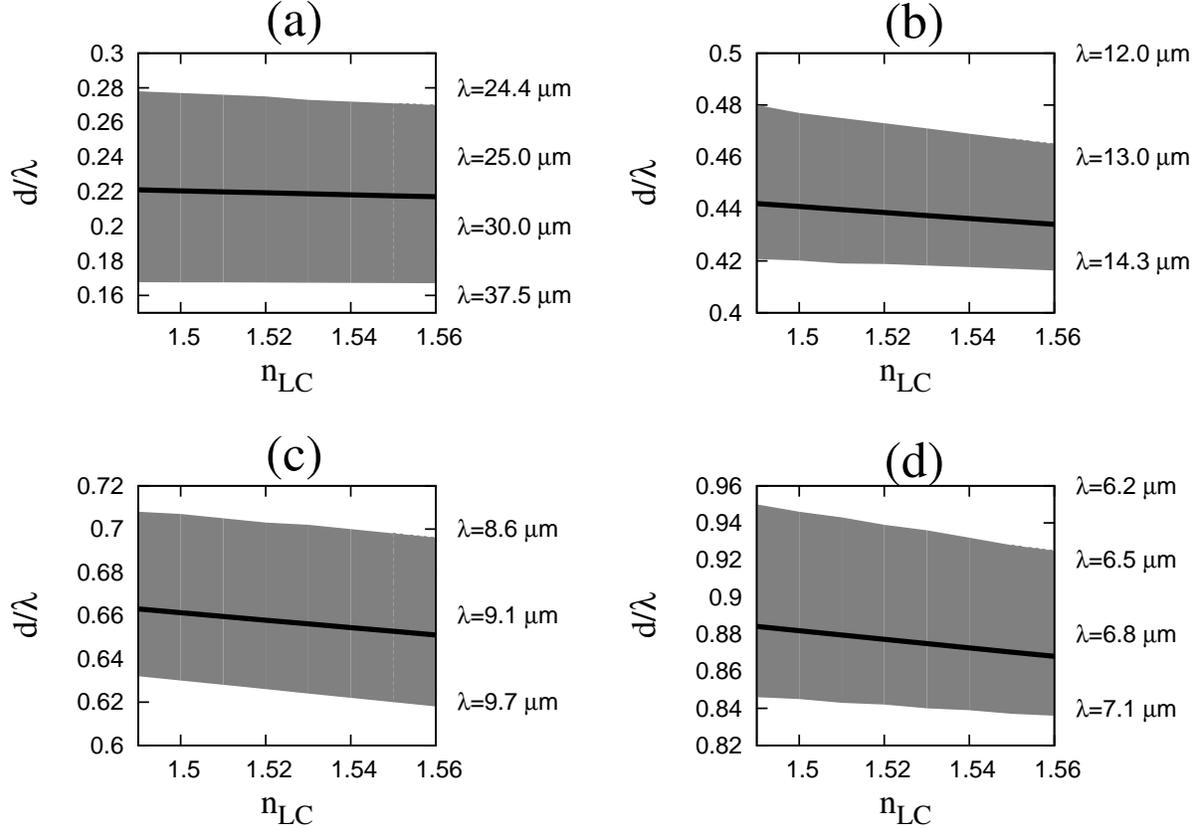}}
\caption{Evolution of the Photonic Band Gap as a function of the variation of the 
liquid crystal refractive index, $n_{LC}$. In panels (a) - (d) we present the variation
for the first four photonic band gaps.}
\end{figure}

\section{Numerical results}

To verify our analytical formula, we have performed a detailed calculation of a 
particular PC. We have chosen the experimental parameters recently reported by
Tolmachev {\it et al.}  in Ref. \cite{Tolmachev_2007a}. 
We consider the PC as  a stack of alternating slabs of dielectric and LC with
 refractive index  $n_{diel}=3.42$ and  $n_{LC}=1.49$, respectively. 
The filling fraction is $f=0.4$. In Fig. 1 we show the Photonic Band Structure (PBS)
calculated with the PWM. \cite{Archuleta_2007} We observe the existence of four band gaps.
We have plot the PBS in the usual way with wave vector ($\pi/d$) and reduced frequency
($d/\lambda$) in the abscissa and ordinate, respectively. Additionally,
in the right side of the ordinate axis, we have plot a wavelength scale. 
This has be done because most of the experimental results are usually
presented in this form in LC experiments. 
In this case, we have illustrated the wavelength scale 
for a period of $d=6 \mu m$ \cite{Tolmachev_2007a}.

We consider a change in $n_{LC}$ as the result of the variation from the homeotropic
 ($n_{LC}=1.49$) to the pseudoisotropic ($n_{LC}=\sqrt{(2n_o^2+n_e^2)/3}=1.56$) 
\cite{Tolmachev_2007a} state under the influence of an external voltage. In Fig. 3
we present the shift of the PBG as the variation of $n_{LC}.$ In panels (a) - (d)
we present  the evolution of the first four PBG, 
which are illustrated in the gray zones. Inside each gray zone, we have plotted the 
evolution of the VLC eq. (7). We observe that for the first PBG in panel (a), the VLC
lies in the middle. For the rest of the cases, the VLC does not defines the 
center of the PBG. However, the interesting fact is that the shift of the PBG's are
predicted. In order to define the stop band shift, we introduce the formula

\begin{equation}
 \Delta \lambda_m = \lambda_{m}(n_{LC}=1.56)-\lambda_{m}(n_{LC}=1.49)
\end{equation}

where $\lambda_{m}(n_{LC}=1.56)$ and $\lambda_{m}(n_{LC}=1.49)$ are
the diffraction for the order $m$. In table 1 we present the 
numerical values obtained with the eqs. (7) and (8).

\begin{table}[h]
\renewcommand{\arraystretch}{1.3}
\caption{Numerical values of the VLC for the first four diffraction orders.}
\vskip0.2in
\begin{center}
\small \begin{tabular}{|c|c|c|c|}\hline
m &  $\lambda_m(n_{LC}=1.49)$ & $\lambda_m(n_{LC}=1.56)$ & $\Delta\lambda_m$   \\ \hline
1  &  $27.144 \mu $m & $ 27.648 \mu$m &  $0.50 \mu$m  \\ \hline
2  &  $13.572 \mu $m & $ 13.824 \mu$m &  $0.25 \mu$m  \\ \hline
3  &  $ 9.048 \mu $m & $  9.216 \mu$m &  $0.16 \mu$m  \\ \hline
4  &  $ 6.786 \mu $m & $  6.912 \mu$m &  $0.12 \mu$m  \\\hline
  \end{tabular}
\end{center}
\label{tab1}
\end{table}

We observe that the greatest shift is given for the diffraction order $m=1$. For the 
following diffraction orders, the shift decreases. It is important to note that for 
the diffraction order $m=3$, we obtain the experimental value obtained in ref. (9).

\section{Conclusion}

In conclusion, we have numerically demonstrated the advantage to use the
VLC to predict the tuning of the stop band in PC structures. 
Previously theorethical reports on tuning were based on calculationd 
on PWE or TMM.  We have shown that our formulation agree well with 
experimental data reported in Ref. (9)
We consider that our formula can be useful to determine in an easy way
the tuning of the stop band in tunable PC.


\end{document}